\documentstyle[11pt]{article}

\title{\bf Cumulative structure function in terms of nucleonic
wave function of the nucleus}
\author{M.A.Braun, \thanks{Permanent address: 
Dep. High-Energy physics,
S.Petersburg University, 198504 S.Petersburg, Russia}
V.M.Suslov \thanks{Permanent address: 
Dep. Math. and Comp, physics,
S.Petersburg University, 198504 S.Petersburg, Russia}
and B.Vlahovic\\
NCCU,Durham, NC, USA}
\date{}
\def\beq{\begin{equation}}
\def\eeq{\end{equation}}
\def\noi{\noindent}
\def\bk{{\bf k}}
\def\tka{\tilde{k}_1}
\def\tkb{\tilde{k}_2}
\def\tr{\tilde{r}}

\begin{document}
\maketitle
\medskip
\vspace{1 cm}

{\bf Abstract}
\vspace{1 cm}

The structure function of the nucleus in the cumulative region $x>1$ is studied
in terms of nucleon degrees of freedom. At high $Q^2$ the resulting expressions
are presented as a sum of contributions from few-nucleon correlations.
Two-nucleon correlations are studied in some detail. Spin variables are averaged out.
In the region $1<x<2$ the  structure functions are calculated for the relativistic 
interaction proposed by F.Gross {\it et al}. They are found to fall with $x$ faster 
than the exponential. For Carbon at $x=1.05$, where the method is not rigorously applicable, 
they turn out to be rougly twice larger than the experimental data. 

\section{Introduction}
The cumulative phenomena, i.e. processes with nuclei
in the kinematical region prohibited for the colliding free nucleons,
are of considerable interest both from the practical and theoretical
poits of view. On the practical side they give a possibility to
effectively raise the energy of the colliding particles,
allowing for subthreshold production of high mass partciles.
On the theoretical side they allow to study small internucleon
distances and thus study the nuclear matter at high densities
and possibly its quark structure. One may speculate that in these
phenomena the cold quark-gluon plasma is formed in the region
of many overlapping nucleons.

Most of the experimental data on the cumulative phenomena refer to
particle production in the  region outside the free nucleon kinematics.
Obtained more than twenty years ago, they were  actively discussed
at that time. Many models were proposed to explain the data.
However no general consensus was reached.
Even the basic question, whether the quark rather than
nucleon degrees of freedom were nesessary to describe the experimental
results remained unsolved. The reason for this lies in the
complexity of the production process. It inevitably involves
soft interaction of the colliding particles,
the description of which is highly  model dependent  and thus
introduces uncontrollable elements into the predictions for cumulative
particle production rates.

From the theoretical point of view a much cleaner observable is the
structure function of the nucleus in the cumulative region, that is,
at the values
of $x$ greater than unity. Unfortunately the relevant experimental
information is
 quite scarce, limited to data on the deuteron [1] and carbon [2]
at relatively low values of $x<1.5$.  However even these modest data present
a clear possibility for theoretical analysis and testing of different
approaches. A description  of the cumulative structure function of the
nucleus  based on the QCD approach and quark structure of
the nucleus was proposed in [3]. Potentially this approach can relate the
observed magnitude and behaviour of the cumulative structure function
to the fundamental QCD parameter $\Lambda$. However the results obtained
in [3] strictly speaking refer to the extreme cumulative kinematical limits.
They also result infrared
unstable and require introduction of an infrared cutoff ("quark mass")
as a parameter. With this single parameter both the slope and the
magnitude of the cumulative structure function were described
reasonably well in [3].

For all that, it remains unclear whether the description of the
nuclear structure function at moderate cumulativity, that is for
$x\sim 2\div 3$ necessarily requires  using the quark language
and thus throwing away most of the information accumulated from the
study of the nucleus in terms of hadrons. It is worth mentioning that
many years ago Frankfurt and Strikman strongly advocated the thesis that
the cumulative phenomena could be well described in terms of 
nucleons, provided the relevant wave functions be appropriately
relativized [4].

In this paper we pursue exactly this approach. Exploiting the fact that
forms for the relativistic  internucleon interaction have lately been
proposed, which lead to excellent agreement with all low energy data,
we use this interaction to calculate the cumulative nuclear structure
function. It is important that the cumulative structure function is
directly related to the  internucleon interaction, unlike most of
other properties which rather involve all sort of averages of the
interaction. Thus our results constitute a stringent test on the
proposed relativistic interaction. In principle our approach allows to
find the cumulative structure function at all values of $x<A$. However
calculational difficulties grow very fast with $x$. For that reason
in this paper, apart from presenting the general formalism, we
calculate the nuclear structure function only in the interval
$1<x<2$ from the contribution which is known as "two-nucleon correlation"
[4], leaving the three-(and more-)nucleon correlations for future studies.

The paper is organized as follows. In Secs.2 and 3 we present our formalism
and separate contributions from 
correlations of several nucleons. In Sec. 4 we analyze the two-nucleon
correlation first for the simplified scalar case and then for the
relaistic spinor case. Sec. 5 presents our nucmerical results.
Some conclusions are drawn in Sec. 6. In the Appendices the
parameters of the used relativistic interaction are presented and
also some calculational details.

\section{Formalism}
At high values of $Q^2$ the
nuclear structure function $F^{(A)}(x,Q^2)$ is given by the ``impulse
approximation'' diagram shown in Fig. 1, corresponding to the
leading twist contribution. Our notations are clear from Fig. 1.
We define in the standard manner $q^2=-Q^2<0$ and
\[x=\frac{Q^2}{2qp}\]
Note that $x$ is defined respective to the collision with a nucleon.
The mass of the nucleus is $M=A(m-\epsilon)$, so that
\[ p^2=m^2-2m\epsilon\equiv m^2-\Delta^2\]
In future we shall neglect the binding energies wherever it is
possible. The physical region for the reaction is evidently
\[(q+Ap)^2\geq M^2\ \  {\rm or}\ \ 0\leq x\leq A\]
All the region
\beq
1\leq x\leq A
\eeq
is cumulative.

The separation of the structure function from the diagram shown in Fig.1
is standard. Let $W_{\mu\nu}^{(A)}$ be  imaginary part of the amplitude
(the discontinuity in $s$ divided by $2i)$. Then one presents
\beq
W_{\mu\nu}^{(A)}=\left(-g_{\mu\nu}+\frac{q_{\mu}q_{\nu}}{q^2}\right)
F^{(A)}_1(x,Q^2)+\frac{1}{qP}\left(P_{\mu}-q_{\mu}\frac{qP}{q^2}\right)
\left(P_{\nu}-q_{\nu}\frac{qP}{q^2}\right)F_2^{(A)}(x,Q^2)
\eeq
with $P=Ap$ the nucleus 4-momentum.
Functions $F_{1,2}^{(A)}$ are the nuclear structure functions.
In the following we concentrate on $F_2^{(A)}$ and suppress  subindex 2.
It is convenient to choose a system in which $q_+=0$. Then taking
the $++$ component of (2) one gets
\beq
W_{++}^{(A)}=\frac{P_+^2}{qP}F^{(A)}(x,Q^2)
\eeq
This relation serves to read the structure function directly from the
diagram shown in Fig. 1.

To move further for simplicity we assume nucleons to be spinless particles.
However the final results will also be valid for realistic nucleons
after averaging over spins. Some
details concerning inclusion of spins will be presented in Section 4.
We present $W_{\mu\nu}^{(A)}$ as an integral over the momentum $k$ of the 
active nucleon
\beq
W_{\mu\nu}^{(A)}=\int\frac{d^4k}{(2\pi)^4}W_{\mu\nu}^{(N)}(q,k)\ 
2\,{\rm Im}\,\Phi(p,k)
\eeq
where $\Phi$ denotes the lower blob in Fig. 1 with the propagators of 
the active nucleon 
included and $W_{\mu\nu}^{(N)}$ is the imaginary part of the upper blob.
The latter can be presented as in (2), with the off-shell structure
functions 
$F_{1,2}^{(N)}(x',Q^2,k^2)$ where $x'=Q^2/(2qk)$. At high $Q^2$ at lower twist
these structure functions do not depend on $k^2$ and coincide with the 
observable
structure functions of the nucleon. At high $Q^2$ in the system $q_+=0$ we have
$q_-\sim Q^2$, $q_{\perp}\sim Q<<q_+$ so that $qp\simeq q_-p_+$ and 
$qk\simeq q_-k_{+}$. So if the scaling variable of the active nucleon is
$z=k_{+}/p_+$, then $x'=x/z$. With these relations we find from (3)
and(4):
\beq
F^{(A)}(x,Q^2)=\frac{1}{A}\int_x^A dz\,
zF^{(N)}\left(\frac{x}{z},Q^2\right) 
\int\frac{p_+dk_{-}d^2k_{\perp}}{(2\pi)^4}
2\,{\rm Im}\,\Phi(p,k)
\eeq
where we used the fact that the structure function of the nucleon depends
only on the scaling variable of the active nucleon.

The integral over $k_{-}, k_{\perp}$ of the Im $\Phi$ can be related to the
distribution in $x$ of the nucleons in the target nucleus.  
To do this consider the baryonic form-factor $B_\mu(q)$ at $q=0$
described by the diagram shown in Fig. 2.
In it the lower blob  is the same 
forward scattering amplitude $\Phi(p,k)$ as in Fig. 1 for the structure
function, only non-cut.
Blob
$\Phi$ is a function of two variables: the  active nucleon virtuality
$m^2-k^2$  and the
square of the c.m. energy of the recoil particles $s=(Ap-k)^2$. 
Neglecting the dependence of 
the nucleon form-factor on the nucleon virtuality we have:
\beq
2P_{\mu}B^{(A)}(0)=\int\frac{d^4k}{(2\pi)^4i}2k_{\mu}
\Phi(p,k)=\int
\frac{dzd^2k_{\perp}ds}{2(2\pi)^4i(A-z)}
2k_{\mu}\Phi(s,k^2)
\eeq
In terms of integration variables in the second integral
\beq
k^2=\frac{z}{A}M_A^2-\frac{z}{A-z}s-\frac{A}{A-z}k_{\perp}^2
\eeq
and the integration over $s$ goes along the Feynman path.

In the complex $s$ plane the singularities of $\Phi$ come, first,
from the standard unitarity righthand cut  and, second, from the
singularities in $k^2$. These lie along the positive values of
$k^2$. However due to a coefficient in (7) they transform into
a lefthand  or righthand cut in $s$ depending on whether the ratio
$z/(A-z)$ is positive or negative. If the cut coming from the
singularities in $k^2$ lies to the right, we shall get zero,
since the integration contour in $s$ then can be closed in the
upper hemisphere with no singularities inside. So, as expected,
 a non-zero result only follows if
\beq 0<z<A \eeq
when the cut from the singularities in $k^2$ lies to the left, and
the result can be obtained by closing the contour around the unitarity
cut. Taking the "+" component of (6) we get
\beq
\int\frac{d^2k_{\perp}}{(2\pi)^2}\int_0^A\frac{zdz}{2\pi(A-z)}
\int_{s_0}^{\infty}\frac{ds}{2\pi}
{\rm Im}\,\Phi(s, k^2)=A^2
\eeq
This relation can be rewritten as a normalization condition
for the relativistic nuclear density in $z$ and $k_\perp$
\beq
\int_0^A\frac{dz}{z}\int\frac {d^2k_\perp}{(2\pi)^2}
\rho(z,k^2_\perp)=1
\eeq
where the density $\rho(z,k_{\perp}^2)$ is defined as an integral over the "-"
component of $k$:
\beq
\rho(z,k_{\perp}^2)=\frac{z^2}{A^2}\frac{1}{2\pi(A-z)}
\int\frac{ds}{2\pi}
{\rm Im}\,\Phi(s, k^2)=\frac{z^2}{A^2}\int \frac{p_+dk_-}{(2\pi)^2}
2\,{\rm Im}\,\Phi(p,k)
\eeq

Note that on the formal level, 
\[
2\,{\rm Im}\,\Phi(s, k^2)=\int d^4r e^{-ikr}
\langle A|N^{\dagger}(r)N(0)
|A\rangle=\]\beq
\sum_{\alpha}(2\pi)^2\delta^4(P_{\alpha}+k-P)
\langle A|N^{\dagger}(0)|A-1,\alpha\rangle\langle A-1,\alpha|N(0)|A\rangle
\eeq
where $|A\rangle$ is the state of the nucleus at rest,
$|A-1,\alpha\rangle$ are the recoiling nuclear states  with $A-1$ nucleons,
including states with $A-1$ free nucleons, 
and $N(x)$ is the
nucleon field operator. The amplitudes $\langle A-1,\alpha|N(0)|A\rangle$
describe splitting of the nucleus into a virtual nucleon and recoil states.
The summation over all
states in (12) is restricted by the energy conservation law
(the conservation of "-" component of the momentum in light-cone
variables). Integration over $k_-$ lifts this restriction and,
as is evident from (12), 
puts the operators $N$ and
$N^{\dagger}$ at equal "time" $x_+=0$. This allows to demonstrate that
in the light-cone formalism with a conserved number of nucleons
$\rho(x,k_{\perp}^2)$ is indeed the relativistic nuclear density
\[
\rho(x,k^2_\perp)=\frac{1}{4\pi}
\int\prod_{j=2}^{A}\frac{dx_jd^2k_{j\perp}}{2x_j(2\pi)^3}
\delta\left(x+\sum_{j=2}^{A}x_j-A\right)
\delta^2\left(k_{\perp}+\sum_{j=2}^{A}k_{j\perp}\right)\]\beq
|\psi(x,k_{\perp};x_2,k_{2\perp};...;x_{A},k_{A,\perp})|^2
\eeq
where $\psi$ is  the standard light-cone wave function of the nucleus,
symmetric in all the nucleons and normalized according to
\beq
\int\prod_{j=1}^{A}\frac{dx_jd^2k_{j\perp}}{2x_j(2\pi)^3}
\delta\left(\sum_{j=1}^Ax_j-A\right)
\delta^2\left(\sum_{j=1}^Ak_{j\perp}\right)
|\psi(x_1,k_{1\perp};x_2,k_{2\perp};...;x_A,k_{A,\perp})|^2=1
\eeq
The derivation is presented in Appendix 1.

The light-cone wave function $\psi$ can be related to the
non-relativistic wave function by taking the non-relativistic limit
in the normalization condition (14), which corresponds to
$x_j\rightarrow 1$. In this limit (14) gives
\beq
\int \prod_{j=2}^A d^3k_j
|\psi(x,k_{\perp};x_2,k_{2\perp};,,,;x_A,k_{A,\perp})|^2=
(16\pi^3)^Am^{A-1}
\eeq
where we introduced $k_{jz}=m(x_j-1)$ for $j>2$. From this we conclude that
in this limit
\beq
\psi(x,k_{\perp};x_2,k_{2\perp};,,,;x_A,k_{A,\perp})\simeq
(16\pi^3)^{A/2}m^{(A-1)/2}\phi(k_1,k_2,...,k_A)
\eeq
where $\phi$ is the non-relativistic wave function of the nucleus at rest.
On the right-hand side $k_j$ denote the 3-dimensional momenta, their sum
equal to zero.

Returning to the structure function  we find from (5)
\beq
F^{(A)}(x,Q^2)=A\int_x^A \frac{dz}{z}\,
F^{(N)}\left(\frac{x}{z},Q^2\right)
\rho(z) 
\eeq
where the distribution in $z$ is just
\beq
\rho(z)=\int\frac{d^2k_{\perp}}{(2\pi)^2}\rho(z,k_{\perp}^2)
\eeq
normalized according to
\beq
\int_0^A\frac{dz}{z}\rho(z)=1
\eeq

\section{Cumulative vs. non-cumulative regions. Correlations}

As indicated, blob $\Phi(p,k)$ includes  two external propagators for 
the active nucleon
and so has a double pole at $k^2=m^2$. If we neglect the binding, then
with all recoil particle at rest we shall have $k=p$ and so $k^2=m^2$, 
which implies
that the distributions $\rho(z,k_\perp^2)$ and $\rho(z)$ are strongly peaked at
$z=1$ and $k_\perp =0$. These values correspond to the non-relativistic
domain, where the density $\rho$ is more or less known from the 
existing nuclear data at relatively small transferred momenta.

Going back to the nuclear structure function we observe from Eq. (17) that
one can distinguish between two different kinematical situations.
If $x\leq 1$ (non-cumulative region) then the integration region 
includes the vicinity of $z=1$, where as we have just seen, $\rho(z)$
has a sharp maximum. Then, as a first approximation, one can take
$F(x',Q^2)$
out of the integral at this point (i.e. at $x'=x$) and integrate the rest
using the normalization condition (19), which gives
\beq
F^{(A)}(x,Q^2)=AF^{(N)}(x,Q^2)
\eeq
One can improve this result by taking some form for  $\rho(z)$ and
integrating over $z$ in a proper way. Then one will get corrections
to (20) (the EMC effect). It is worth noting that  it is
always possible to match the experimental data by appropriately choosing
$\rho$ at finite (relativistic) values of $|z-1|$.

However in the cumulative region $x>1$ the point $z=1$ stays outside
the integration region in (17). This means that in this interval of $x$
values of $\rho$ are important which come from $\Phi(p,k)$ with
the active nucleon far off-shell, well outside the non-relativistic region.
From the point of view of the latter, it is the far relativistic
asymptotics of the nuclear dynamics which now matters.

The technique to study this asymptotics is in fact standard.
Let us consider a simple example of the deuteron, for which $\Phi(p.k)$ reduces 
to the square of the Bethe-Salpiter wave function $\psi^{(d)}(k,2p-k)$
satisfying the equation
\beq
(m^2-k_1^2)(m^2-k_2^2)\psi^{(d)}(k_1,k_2)=
\int\frac{d^2k'_1}{(2\pi)^4}V(k_1,k_2;k'_1,k'_2)\psi^{(d)}(k'_1,k'_2)
\eeq
where $k_1+k_2=k'_1+k'_2=2p$. In the vicinity of the mass shell 
$k_1^2=k_2^2=m^2$ this equation reduces to the ordinary Schroedinger
equation and correspondingly $\psi^{(d)}$ reduces to its nonrelativistic
limit $\psi^{(d)}_{NR}$. Now suppose we want to study the asymptotics of
$\psi^{(d)}$
far off shell. A simple way do to it is to rewrite (22) as
\beq
\psi^{(d)}(k_1,k_2)=\frac{1}{(m^2-k_1^2)(m^2-k_2^2)}
\int\frac{d^2k'_1}{(2\pi)^4}V(k_1,k_2;k'_1,k'_2)\psi^{(d)}(k'_1,k'_2)
\eeq
and take into account that the integration is  dominated by the
close to on-shell values of the momenta, that is, by the non-relativistic
region. So inside the integral one can substitute $\psi^{(d)}$ by its 
non-relativistic limit and moreover take the interaction $V$ outside
the integral at small values of the 3-momenta $k'_{1,2}$ corresponding
to the nonrelativistic region. In this way one gets the desired asymptotics
as
\beq
\psi^{(d)}(k_1,k_2)\sim\frac{1}{(m^2-k_1^2)(m^2-k_2^2)}
V(k_1,k_2;p,p)
\int\frac{d^2k'_1}{(2\pi)^4}\psi_{NR}^{(d)}(k'_1,k'_2)
\eeq
where $p=(m-\epsilon,\vec{0})$. In terms of diagrams this 
procedure corresponds to 
transforming the NND vertex as shown in Fig. 3, where it is assumed that the 
internal lines correspond to the non-relativistic deuteron and the 
external ones correspond to the relativistic deuteron.

For a nucleus consisting of more than two nucleons this procedure
leads to the diagrams shown in Fig. 4$a$. If more than two nucleons
are relativistic, it can be repeated, leading to diagrams like Fig. 4$b$.
Following [4] one speaks of two- (Fig. 4$a$) or three- (Fig. 4$b$) or
more-body correlations in the nucleus to describe two, three or more
relativistic nucleons. 

From the pure kinematical considerations one immediately concludes that
the $k$-fold correlation gives a contribution to the distribution $\rho(x)$
in the region
\[0\leq x\leq k\]
Equivalently, choosing a particular interval of $x$
\beq
n-1\leq x\leq n
\eeq
the contribution to the nuclear structure function in this region
comes from correlations of $k$ nucleons with $k\geq n$.

At large virtualities, with interaction and propagators falling like powers,
one finds that the contribution from the $k$-fold correlations
diminishes fast with the growth of $k$. The bulk of the contribution
in the region (24) will then come from precisely the $n$-fold correlation,
the ones with $k>n$ giving only a small correction.
So, in the first approximation, to study the structure function in the
region (24) one has to take into account only the $n$-fold correlation.

Up to now the cumulative nuclear structure function has 
been studied experimentally only in the region
\beq
1<x<2
\eeq
This gives motivation to start from the pair correlations in the nucleus,
which will be the subject of the next sections. 

\section{Two-body correlations}
\subsection{Scalar nucleons}
To clearly formulate our approach we continue with the scalar
nucleons case.
As explained in the previous section, the contribution of the 2-body
correlations to function $\Phi(p.k_1)$ corresponds to the diagram
shown in Fig. 5 in which high- and low-momenum particles are indicated
with thick 
and thin lines respectively. 
 Explicitly the contribution of the diagram in Fig. 5 can be written as
\[
2\,{\rm Im}\,\Phi(p,k_1)=\frac{1}{(m^2-k_1^2)^2}
\int\prod_{j=1}^2\frac{d^4k'_jd^4k''_j}{(2\pi)^8}
(2\pi)^4\delta^4(k'_1+k'_2-k''_1-k''_2)\]\beq
2\pi\delta(k_2^2-m^2)
V(k_1,k_2|k'_1,k'_2)V(k_1,k_2|k''_1,k''_2)2\,{\rm Im}\,\Phi_2(p;k'_1,k'_2;
k''_1,k''_2)
\eeq
Here $V$ is the relativistic (Bethe-Salpeter) potential and $\Phi_2$
corresponds to the lower blob with all intermediate states of $A-2$
nucleons.

The high-momentum part includes the first factor, potentials and 
the $\delta$ -function coming from the real nucleon number 2.
We can take the latter factors  out of the integral at 
the point which corresponds to
the initial nucleons at rest, that is at $k'_j=k''_j=p$, $j=1,2$ and
$p=(m-\epsilon,\vec{0})$. Then the high-momentum part separates as a factor, so that we
get
\beq
2\,{\rm Im}\,\Phi(p,k_1)=  
2\pi\delta((k_1-2p)^2-m^2)A^2(A-1)w_2H(p,k_1).
\eeq
Here the high-momentum factor $H$ is
\beq
H(p,k_1)=H(k_1^2)=\frac{V^2(k_1,2p-k_1|p,p)}{(m^2-k_1^2)^2}
\eeq
($H$ depends only on $k_1^2$ since the product $pk_1$ can be related to
it from the condition $k_2^2=(2p-k_1)^2=m^2$).
Factor $w_2$ is just a constant which accumulates all the information from
the nucleus:
\beq
A^2(A-1)w_2=
\int\prod_{j=1}^2\frac{d^4k'_jd^4k''_j}{(2\pi)^8}
(2\pi)^4\delta^4(k'_1+k'_2-k''_1-k''_2)2\,{\rm Im}\,\Phi_2(p;k'_1,k'_2;
k''_1,k''_2)
\eeq

The calculation of $w_2$, is more or less standard. 
Similarly to ${\rm Im}\,\Phi$, the imaginary part of $\Phi_2$ can be 
presented in the form
\[
2\,{\rm Im}\Phi_2(p;k_1,k_2;\tka\tkb)
=\int d^4R d^4r d^4\tr e^{iRK+ir(k_1-K/2)-i\tr(\tka-K/2)}
\]\beq
\langle A|T\{N^{\dagger}(R+r/2)N^{\dagger}(R-r/2)\}
T\{N(\tr/2)N(-\tr/2)\}|A\rangle
\eeq
where $K=k_1+k_2=\tka+\tkb$ and, as before, $N (N^{\dagger})$ are the 
annihilation (creation) operators
of the nucleon field and $|A\rangle$ is the nucleus state at rest.
Putting this representation into (29) we find that the mometum integration
puts all coordintes to zero:
\beq
A^2(A-1)w_2=\langle A|(N^{\dagger}(0))^2N^2(0)|A\rangle
\eeq
To relate this expression with the standardly defined quantities, we note that
all particles may be considered non-relativistic. So we may pass to 
non-relativistic states and operators. 
\beq
|A\rangle=\sqrt{2Am}|A_{NR}\rangle,\ \ N(N^{\dagger}(r)=\frac{1}{\sqrt{2m}}
N_{NR}(N^{\dagger}_{NR})(r)
\eeq
Thus in terms of non-relativistic quantities we get (suppressing subindex NR)
\beq
A^2(A-1)w_2=\frac{A}{2m}\langle A|(N^{\dagger}(0))^2N^2(0)|\rangle=
\frac{A^2(A-1)}{2m}\int\prod_{i=3}^Ad^3r_i|\psi^{(A)}(0,0,{\bf r_3,...r_A})|^2
\eeq
where $\psi^{(A)}$ is the non-relativistic wave function of the target nucleus.
It appears at zero distance between the 1st and 2nd nucleons, 
which is physically natural
and corresponds to a correlation between them. To express (33) via better
known quantities,
one may take into account the translational invariance by presenting
\beq
|\psi^{(A)}({\bf r_1,r_2, r_3,...r_A})|^2=
\int d^3r\rho^{(A)}({\bf r_1+r,r_2+r,...r_A+r})
\eeq
with the nuclear $\rho$-matrix normalized as
\beq
\int\prod_{i=1}^Ad^3r_i\rho^{(A)}({\bf r_1,...r_A})=1
\eeq
In terms of this $\rho$-matrix we find from (33)
\beq
w_2=\frac{1}{2m}\int d^3r \rho^{(A)}({\bf r, r})
\eeq
where 
\beq
\rho({\bf r_1,r_2})=\int\prod_{i=3}^Ad^3r_i\rho^{(A)}({\bf
r_1,r_2,r_3,....r_A})
\eeq
and is normalized to unity after integration over ${\bf r_1}$ and ${\bf r_2}$.
The form (34) admits presenting $\rho^{(A)}$ as a product of
single-nucleon distributions. In this approximation
\beq
w_2=\frac{1}{2m}\int d^3r (\rho^{(A)}({\bf r}))^2
\eeq
and thus is directly expressed via the nuclear density.

\subsection{Distributions in $x$ and $k_{\perp}^2$}

According to (11), to find the distributions in $z$ and $k_{\perp}^2$ we
have to integrate the
obtained function $\Phi(p,k)$ over $k_-$. This integration is trivial due
to the $\delta$-function in (27). It fixes the value of $k^2$ to be
\beq
k^2=zm^2\frac{3-2z}{2-z}-k_{\perp}^2\frac{2}{2-z}-\Delta^2\frac{2z}{2-z}
\eeq
One finds that $\rho(z,k_\perp^2)$ is different from zero in the region
(see Appendix 2)
\beq
k_\perp^2\geq\Delta^2\left(2-z-\frac{2z}{A-2}\right)-m^2(z-1)^2
\eeq
This condition is operative only if we take into acount the binding
energy $\Delta^2/m$. Otherwise it is satisfied at all $k_{\perp}$ and $1<z<2$.

The internucleon relativitic potential can be taken
as a sum of one-boson exchange  contributions
\beq
V(k_1,k_1|k'_1,k'_2)=\sum_i\frac{g^2_i}{\mu_i^2-q^2}
\eeq 
where $q=k_1-k'_1$ is the momentum transfer.
In our case $k'_1=k'_2=p$, $k_1=k$ and $k_2=2p-k$. So one easily finds
\beq
q^2=\frac{1}{2}(k_1^2-m^2)+\Delta^2
\eeq

The distribution in $z$ and $k_\perp^2$ is obtained directly from (11):.
\beq
\rho(z,k_\perp^2)=w_2\frac{z}{4\pi(2-z)}H(k^2)=
w_2\frac{z}{4\pi(2-z)}
\frac{1}{(m^2-k^2)^2}
\sum_{il}\frac{g^2_ig^2_l}{(\mu_i^2-q^2)(\mu_l^2-q^2)}
\eeq
where $k^2$ and $q^2$ are expressed via $z$ and $k_\perp^2$ by Eqs. (39)
and (42) and $w_2$ is given by (38).

The remaining task is to integrate over $k_\perp^2$ to obtain the final
distribution in $z$, which  is straightforward. Rather than discuss it
we proceed to the realistic spinor case.

\subsection{Spinor nucleons}
A rigorous treatment of the spinor nucleon case requries knowledge
both of the spinor structure of the virtual photon-nucleon amplitude and
the nucleus wave function, which is far beyond present (and future)
possibilities. For this reason it is natural to  use spin-averaged
quantities for both quantities. More concretely we separate the
integration over the virtual nucleon momentum and present the
contribution from the diagram in Fig. 1 (right-hand side of Eq. (4)) in
the form
\beq
D=\int d^4k
{\rm Sp}\{(m+\hat{k})U(q,k)(m+\hat{k})L(p,k)\}
\eeq
where $U(p,k)$ is the upper blob (virtual $\gamma$-N amplitude) and
$L(p,k)$ is the lower blob (with factor $1/(m^2-k_1^2)^2$ included).
Our approximation is then to take
\beq
{\rm Sp}\{(m+\hat{k})U(q,k)(m+\hat{k})L(p,k)\}\simeq
\frac{1}{2}{\rm Sp}\{(m+\hat{k})U(p,k)\}
{\rm Sp}\{(m+\hat{k})L(p,k)\}
\eeq
which amounts to taking spin averaged values multiplied by the number
of the spin states (two).
After this averaging is made, all our formulas derived for the scalar
case remain valid also for spinor nucleons.

Let us apply this recipe for the contribution from the two-nucleon
correlation, corresponding to Fig. 5.
In the spinor case instead of Eq. (26) we find for the
function $2\,{\rm Im}\,\Phi(p,k_1)$:
\[
2\,{\rm Im}\,\Phi(p,k_1)=\frac{1}{(m^2-k_1^2)^2}
\int\prod_{j=1}^2\frac{d^4k'_jd^4k''_j}{(2\pi)^8}
(2\pi)^4\delta^4(k'_1+k'_2-k''_1-k''_2)
2\pi\delta(k_2^2-m^2)\]\beq
{\rm Sp}\,\Big\{2\,{\rm Im}\,\Phi_2(p;k'_1,k'_2;k''_1,k''_2)
(m+\hat{k}'_1)_1(m+\hat{k}'_2)_2C(k_1,k_2|k'_1,k'_2|k''_1,k''_2)
(m+\hat{k}''_1)_1(m+\hat{k}''_2)_2\Big\}
\eeq
where $C$ is the two-nucleon correlation blob:
\beq
C(k_1,k_2|k'_1,k'_2|k''_1,k''_2)=V(k'_1,k'_2|k_1,k_2)(m+\hat{k}_1)
(m+\hat{k}_2)V(k_1,k_2|k''_1,k''_2)
\eeq
and $V$ is the relativistic interaction. Both $C$ and $V$ are matrices in
in spinor indeces of both nucleons. Operators
$(m+\hat{k})_{1(2)}$ act on the spinor indeces of 1st (2nd) nucleon.
The trace is to be taken over spinor indeces of both nucleons.
In deriving (46)
we have applied (45) to
average over the spin states of the active nucleon (momentum $k_1$). Note
that the factor 1/2 is to be included in the definition of the spin-
averaged nucleon structure function $F$ in Eq. (5).

Now we use the fact that the momenta $k'_1,k_2,k''_1$ and $k''_2$
are non-relativistic and also  apply the
recipe (45) two more times to sum over spin states of
the nucleons 1 and 2 before the left interaction and after the right
interaction in Fig. 5 to obtain a representation identical to (27)
where now
\beq
H(k_1^2)=\frac{1}{4}
{\rm Sp}\{(m+\hat{p})_1(m+\hat{p})_2C(k_1,2p-k_1|p,p|p,p)\},
\eeq
and
\[
A^2(A-1)w_2=
\int\prod_{j=1}^2\frac{d^4k'_jd^4k''_j}{(2\pi)^8}
(2\pi)^4\delta^4(k'_1+k'_2-k''_1-k''_2)\]\beq
{\rm Sp}\,\Big\{2\,{\rm Im}\,\Phi_2(p;k'_1,k'_2;k''_1,k''_2)
(m+\hat{p})_1(m+\hat{p})_2\Big\}
\eeq

In the latter expression in the non-relativistic approximation we may write
\[
{\rm Sp}\,\Big\{2\,{\rm Im}\,\Phi_2(p;k'_1,k'_2;k''_1,k''_2)
(m+\hat{p})_1(m+\hat{p})_2\Big\}\]\[\simeq
\frac{1}{4m^2}{\rm Sp}\,\Big\{(m+\hat{k}'_1)(m+\hat{k}'_2)\,2\,{\rm Im}\,
\Phi_2(p;k'_1,k'_2;k''_1,k''_2)
(m+\hat{k}''_1)(m+\hat{k}''_2)\Big\}\]
\[
=\int d^4R d^4r d^4\tr e^{iRK+ir(k'_1-K/2)-i\tr(k''_1-K/2)}
\]\beq
{\rm Sp}\,\langle A|T\{N^{\dagger}(R+r/2)N^{\dagger}(R-r/2)\}
T\{N(\tr/2)N(-\tr/2)\}|A\rangle
\eeq
where now $N(N^{\dagger})$ are 4-spinor operators of the nucleon field.
In the non-relativistic limit their first two components
 reduce to non-relativistic spinor
fields for the annihilation (creation) of a nucleon with a given spin component 
at a given point, the second two reducing to zero.  
As in (30), the integration over the momenta in (50) puts all coordinates
to zero
and we obtain similarly to (33)
\beq
A^2(A-1)w_2=\frac{A}{2m}\sum_{\sigma,\sigma_2=1}^2
\langle A|(N^{\dagger}_{\sigma_1}(0)N^{\dagger}_{\sigma_2}(0)
N_{\sigma_1}(0)N_{\sigma_2}(0)|\rangle \]\[ =
\frac{A^2(A-1)}{2m}\int\prod_{i=3}^Ad^3r_i\prod_{j=1}^A\sum_{\sigma_j=1}^2
|\psi^{(A)}_{\sigma_1,\sigma_2,...\sigma_A}(0,0,{\bf r_3,...r_A})|^2
\eeq
where $\sigma$'s denote non-relativistic spin projections of the nucleons.
Obviously the integral on the right-hand side has the same meaning as in the 
scalar case: the total probability to have the two active nucleons at the origin
irrespective of their spins. Presenting the spin-summed square modulus of the
nuclear wave function in the form (34) one arrives at the same formulas
(36) or (38) as for the scalar case.

So in the end the only difference from the scalar case is
the expression (48) for the
function $H$ as a spin average over the two-nucleon correlation
contribution given by Eq. (47) in terms of the relativistic internucleon
interaction.

\section{Numerical results}
Calculation of the cumulative nuclear structure function clearly splits
into three steps. First the density $\rho(x,\bk)$ has to be determined
from the given relativistic internucleon potential $V$. Second this
density has to be intergrated over the transverse momenta to find the
resulting distribution in $x$. Finally one should convolute this
distribution with the known structure function of the nucleon.

To determine $\rho(x,\bk)$ one first should calculate the
two-nucleon correlation factor $H$ according to Eqs. (47) and (48).
We borrow the relativistic internucleon potential $V$ from [5],
where it is represented by a sum of $\pi$, $\eta$, $\sigma$, $\omega$
and $\rho$ exhanges, modified by form-factors to cut the high momentum
contribution and adapted to off-shell nucleons. In fact four different
parametrizations of the potential are presented in [5]. They are given
in Appendix 1 together with the concrete form of various contributions.
As we shall see different parametrizations give practically
the same results for the cumulative densities and structure functions.

The calculation of $H$
is rather tedious due to many different terms and relativistic
traces. Some intermediate formulas are given in Appendix 2.

The nuclear factor $w_2$ was calculated from the standard Woods-Saxon
nuclear density. Obviously it is of the order $1/A$ and has a dimension
of mass squared. Correspondingly in Fig. 6 we show its
values multiplied by $(A-1)/m^2$
for different nuclei. The resulting values  steadily grow
with $A$. Their $A$-dependence does not correspond to  $A^{\alpha}$:
one finds that in this parametrization $\alpha$ goes down from
0.3 to 0.15 as $A$ rises from 20 to 200.

Intergrating the found $\rho(x,\bk)$ over the transverse momenta we found
the distribution $\rho(x)$. In Fig. 7 we show its values for Carbon with two
potential models from [5]: IA and IIA. One observes that up to $x\sim
1.3$ both potentials gives the same $\rho(x)$. At higher $x$ model IA
gives  values for $\rho(x)$ higher by factor $1.5\div 3$, which is quite
a small difference on the cumulative scale.

The resulting
cumulative structure function of Carbon are presented in
Figs. 8 and 9 at $Q^2=3,30$ and 1000 (Gev/c)$^2$ for potental models IA
and IIA respectively. 
 For the nucleon
structure function
$F(x,Q^2)$ the GRV LO parametrization has been taken from [6].
The results are quite similar for the two potential models,
both in form and magnitude. 
One observes that the cumulative structure function turns out to be
rather weakly dependent on  $Q^2$: at moderate $x$ it falls with
$Q^2$ by a factor of
the order 4, as $Q^2$ rises from 3 to 1000 (GeV/c)$^2$.
As to the $x$ dependence it is evidently non-exponential, although
universal in $Q^2$, the slope of
$F^{(A)}(x)$ changing from 10 to 40 as $x$ rises from 1.2 to 1.7.

Comparison with the experimental data from [2] is presented in Table 1.
One observes that at $x=1.05$ and 1.15 our results lie above the data
by roughly 100\% and 50\% respectively. This disagreement is not so
dramatic considering the steep fall of the structure function
beyond $x=1$. One has also take into account that these values of $x$
are too close to unity for our formalism to work well.

\begin{center}
{\large\bf Table 1. Structure Function of Carbon (x 10$^6$)\\
in different potential models of [5]}\vspace{0.8 cm} 

\begin{tabular}{|r|c|r|r|r|r|r|}
\hline
 $x$   &$Q^2$ (GeV/c)$^2$ & IA& IB& IIa & IIB& Exp\\\hline
1.05&   61.& 162.  & 153.&    126.&    153.&64.1$\pm$7.4\\
    &   85.& 150.  & 142.&    116.&    142.&47.8$\pm$5.2\\
    &  150.& 133.  & 126.&    103.&    125.&45.8$\pm$5.1\\\hline         
1.15&   61.&  39.5 &  40.7&   30.9&     39.7&$<$27.0\\
    &   85.&  36.4 &  37.6&   28.5&     36.7&$<$15.1\\
    &  150.&  32.0 &  33.1&   25.0&     32.3&$<$22.3\\\hline  
1.30&   61.&   3.65&   4.15&  2.71&      3.89&$<$6.9\\
    &   85.&   3.35&   3.81&  2.48&      3.56&$<$5.4\\
    &  150.&   2.90&   3.30&  2.15 &     3.09&$<$3.8\\\hline
\end{tabular}
\end{center}

\section{Conclusions}

We have developed a method which allows to calculate the nuclear structure
function at $x>1$ in terms of nucleonic degrees of freedom with full
relativistic kinematics. The method avoids solving the complete
relativistic many-body problem and relates the structure function directly
to the high-momentum asymptotics of the relativistic potentials.
While for the 2-body correlations it can be considered as a simplification
of the solution of the Bethe-Salpeter or Gross equations, for the
correlations involving more nucleons it seems to be the only realistic
approach.

The results of our calculations confirm that the nuclear structure
function at $x>1$ depends on $Q^2$ only weakly. Its $x$-behaviour
cannot be well desribed by an exponential. Comparison to (scarce)
experimental data for C at $x=1.05$ and 1.15 shows that  our results
have the same order of magnitude, lying above the experimental data by
100\% and 50\% respectively. This disagreement may be completely explained
by the comparatively low values of $x$ where our formalism is not
supposed to work well. One should wait for more data at higher $x$ to draw
final conclusions as to the validity of the chosen picture and the
potentials.

 \section{Acknowledgements}
\vskip.15in
M.A. Braun and V.M.Suslov are deeply thankful to the NCCU, NC,  USA
for the hospitality and financial support.

\section{References}

1. W.P.Shuetz {\it et al. }, Phys. Rev. Lett. {\bf 38} (1977) 259.\\
2. A.C.Benvenuti {\it et al.}, BCDMS Collab., Z.Phys. {\bf C63} (1994) 29.\\
3. M.Braun and V.V.Vechernin, Nucl. Phys. {\bf B427} (1994) 614.\\
4. L.Frankfurt and M.Strikman, Phys.Rep.{\bf 76} (1981) 215.\\
5. F.Gross, J.W.Van Orden and K.Holinde, Phys. Rev C45 (1992) 2094-2132. \\
6. M.Glueck, E.Reya and A.Vogt, Z.Phys. {\bf C67} (1995) 433.\\

\section{Figure captions}

\noi Fig. 1. The impulse approximation for the nuclear structure function.\\
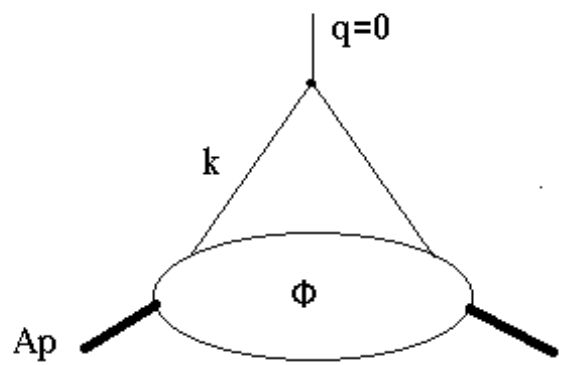
Fig. 2. The impulse approximation for the nuclear form-factor.\\
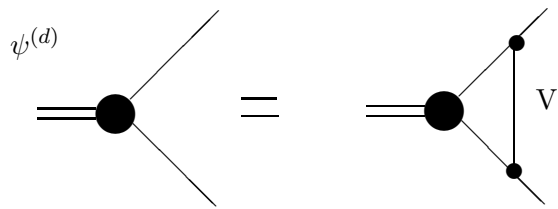
Fig. 3. The asymptotics of the Bethe-Salpeter function from an iteration
of the acting potential.\\
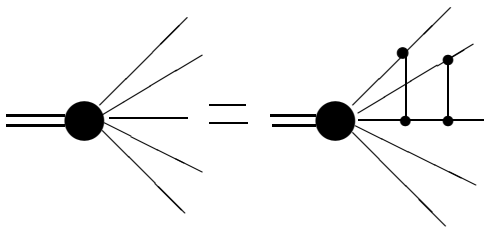
Fig. 4. The two-nucleon (a) and three-nucleon (b) correlations in
the nucleus.\\
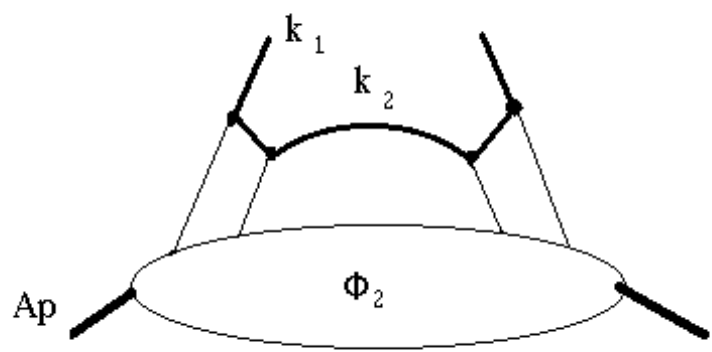
Fig. 5. The contribution from the two-nucleon correlation
(Eqs. (26) and (46)).\\
Fig. 6. The nuclear factor $(A-1)w_2/m^2$ as a function of $A$ .\\
Fig. 7. The distribution $\rho(x)$ for $^{12}$C in the cumulative region
for potential models IA (upper curve) and IIA.\\
Fig. 8. The cumulative structure function of $^{12}$C with the potetial
model IA.
Curves from top to bottom correspond to $Q^2=3,10,30,100$ and
1000 (GeV/c)$^2$.\\
Fig. 9. Same as Fig. 8 with a potential model IIA.\\

\newpage
\unitlength=1mm
\special{em:linewidth 0.4pt}
\linethickness{0.4pt}
\begin{picture}(132.67,137.00)
\put(20.00,123.00){\circle*{5.20}}
\put(9.67,123.67){\line(1,0){7.33}}
\put(17.00,123.67){\line(1,0){0.33}}
\put(9.67,122.33){\line(1,0){7.67}}
\put(22.00,125.00){\line(1,1){11.67}}
\put(22.33,121.67){\line(1,-1){11.00}}
\put(36.67,125.00){\line(1,0){4.67}}
\put(36.67,122.67){\line(1,0){5.00}}
\put(63.67,123.33){\circle*{5.20}}
\put(53.33,124.00){\line(1,0){7.33}}
\put(60.67,124.00){\line(1,0){0.33}}
\put(53.33,122.67){\line(1,0){7.67}}
\put(65.67,125.33){\line(1,1){11.67}}
\put(66.00,122.00){\line(1,-1){11.00}}
\put(73.00,132.67){\line(0,-1){18.00}}
\put(73.67,131.67){\circle*{0.00}}
\put(73.33,132.33){\circle*{2.00}}
\put(73.00,115.33){\circle*{2.00}}
\put(9.33,132.33){\makebox(0,0)[cc]{$\psi^{(d)}$}}
\put(77.33,125.00){\makebox(0,0)[cc]{V}}
\put(107.67,125.67){\makebox(0,0)[cc]{Fig. 3}}
\end{picture}
\newpage
\begin{picture}(132.67,137.00)
\put(19.00,84.00){\circle*{5.20}}
\put(52.33,84.00){\circle*{5.20}}
\put(86.00,84.33){\circle*{5.20}}
\put(115.67,84.67){\circle{0.94}}
\put(115.67,84.67){\circle*{5.20}}
\put(8.67,84.67){\line(1,0){7.33}}
\put(16.00,84.67){\line(1,0){0.33}}
\put(8.67,83.33){\line(1,0){7.67}}
\put(43.67,84.67){\line(1,0){6.00}}
\put(44.00,83.33){\line(1,0){5.67}}
\put(77.00,84.67){\line(1,0){6.00}}
\put(77.33,83.67){\line(1,0){6.00}}
\put(107.67,85.00){\line(1,0){5.00}}
\put(107.67,83.33){\line(1,0){5.33}}
\put(22.33,84.33){\line(1,0){10.33}}
\put(55.33,84.00){\line(1,0){12.00}}
\put(89.00,84.33){\line(1,0){11.00}}
\put(118.67,84.33){\line(1,0){13.67}}
\put(21.00,86.00){\line(1,1){11.67}}
\put(21.33,82.67){\line(1,-1){11.00}}
\put(54.67,86.00){\line(1,1){13.00}}
\put(54.67,82.33){\line(1,-1){12.33}}
\put(87.67,86.00){\line(1,1){13.00}}
\put(88.00,82.67){\line(1,-1){12.33}}
\put(117.67,86.00){\line(1,1){14.33}}
\put(117.67,82.67){\line(1,-1){15.00}}
\put(61.67,93.00){\line(0,-1){9.00}}
\put(94.67,93.33){\line(0,-1){17.33}}
\put(126.00,84.33){\line(0,-1){10.33}}
\put(61.33,93.00){\circle{1.49}}
\put(61.33,93.00){\circle*{1.49}}
\put(61.67,84.00){\circle*{1.33}}
\put(94.67,92.67){\circle*{1.33}}
\put(94.67,76.33){\circle*{1.49}}
\put(126.00,84.33){\circle*{1.49}}
\put(126.00,74.33){\circle*{1.49}}
\put(35.67,86.00){\line(1,0){4.67}}
\put(69.67,84.33){\line(1,0){4.33}}
\put(71.67,86.67){\line(0,-1){5.00}}
\put(102.33,84.33){\line(1,0){3.33}}
\put(104.00,86.33){\line(0,-1){4.33}}
\put(35.67,83.67){\line(1,0){5.00}}
\put(19.33,38.00){\circle*{5.20}}
\put(52.67,38.00){\circle*{5.20}}
\put(9.00,38.67){\line(1,0){7.33}}
\put(16.33,38.67){\line(1,0){0.33}}
\put(9.00,37.33){\line(1,0){7.67}}
\put(44.00,38.67){\line(1,0){6.00}}
\put(44.33,37.33){\line(1,0){5.67}}
\put(22.67,38.33){\line(1,0){10.33}}
\put(55.67,38.00){\line(1,0){12.00}}
\put(21.33,40.00){\line(1,1){11.67}}
\put(21.67,36.67){\line(1,-1){11.00}}
\put(55.00,40.00){\line(1,1){13.00}}
\put(55.00,36.33){\line(1,-1){12.33}}
\put(62.00,47.00){\line(0,-1){9.00}}
\put(61.67,47.00){\circle{1.49}}
\put(61.67,47.00){\circle*{1.49}}
\put(62.00,38.00){\circle*{1.33}}
\put(36.00,40.00){\line(1,0){4.67}}
\put(36.00,37.67){\line(1,0){5.00}}
\put(21.00,38.33){\line(5,3){14.00}}
\put(22.00,37.67){\line(2,-1){13.00}}
\put(55.67,39.00){\line(5,3){15.33}}
\put(55.33,37.33){\line(2,-1){16.00}}
\put(74.67,40.00){\line(0,1){0.33}}
\put(66.33,38.00){\line(1,0){6.00}}
\put(66.33,38.00){\line(1,0){1.33}}
\put(67.67,46.00){\line(0,-1){8.00}}
\put(67.67,46.00){\circle*{1.33}}
\put(67.67,38.00){\circle*{1.33}}
\put(69.67,61.00){\makebox(0,0)[cc]{a}}
\put(69.33,13.00){\makebox(0,0)[cc]{b}}
\put(107.67,34.67){\makebox(0,0)[cc]{Fig. 4}}
\end{picture}
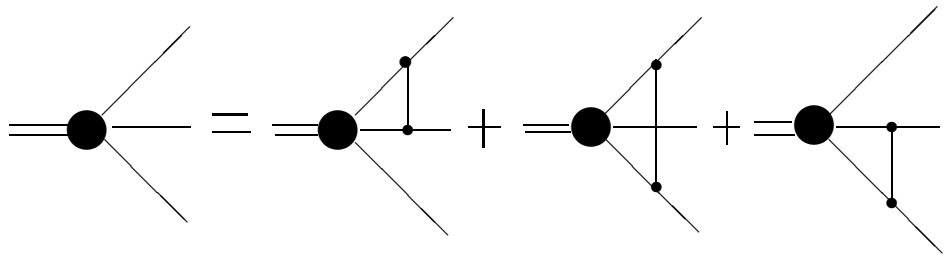
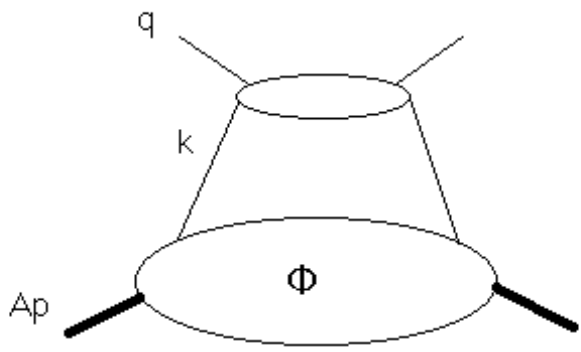
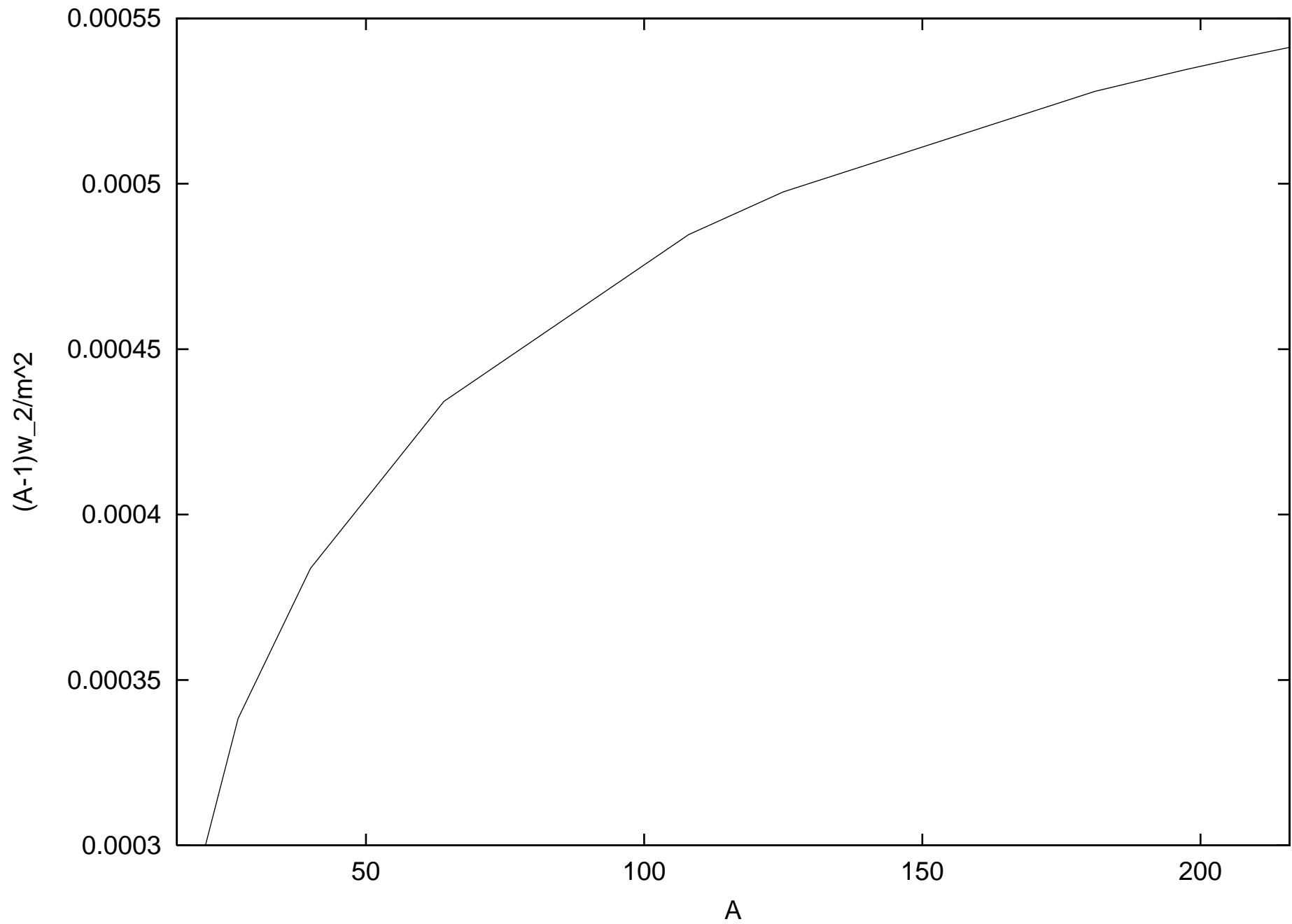
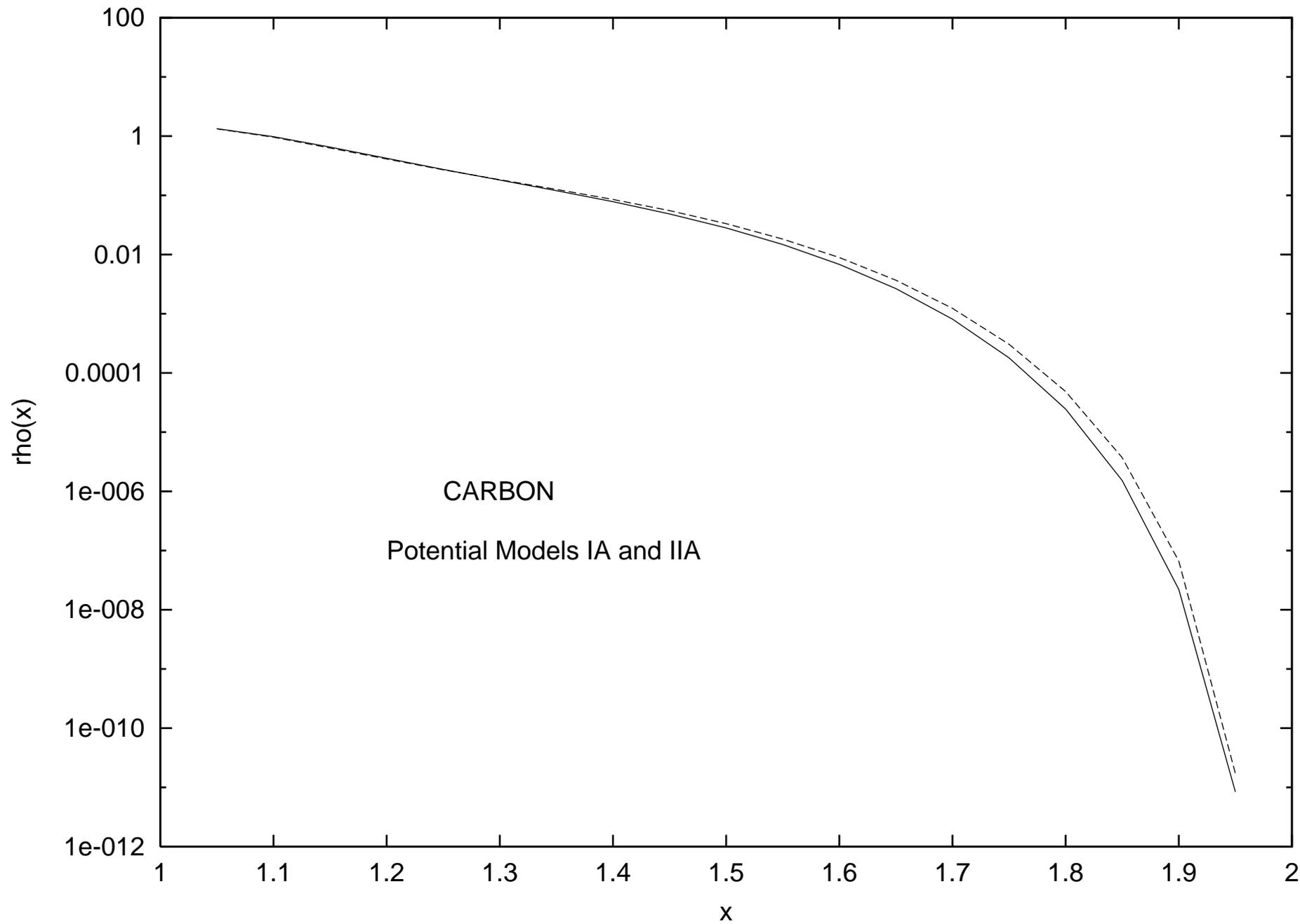
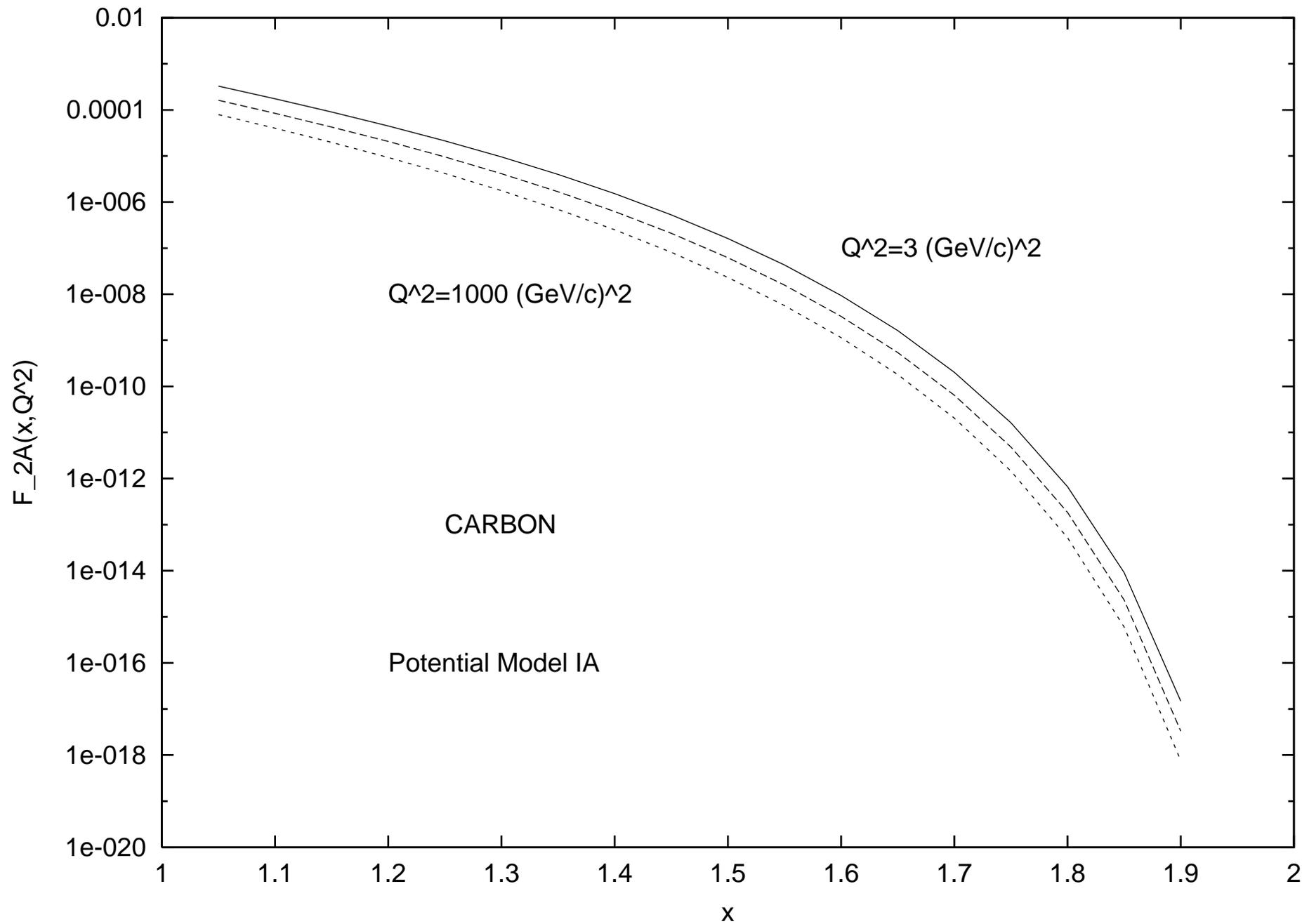
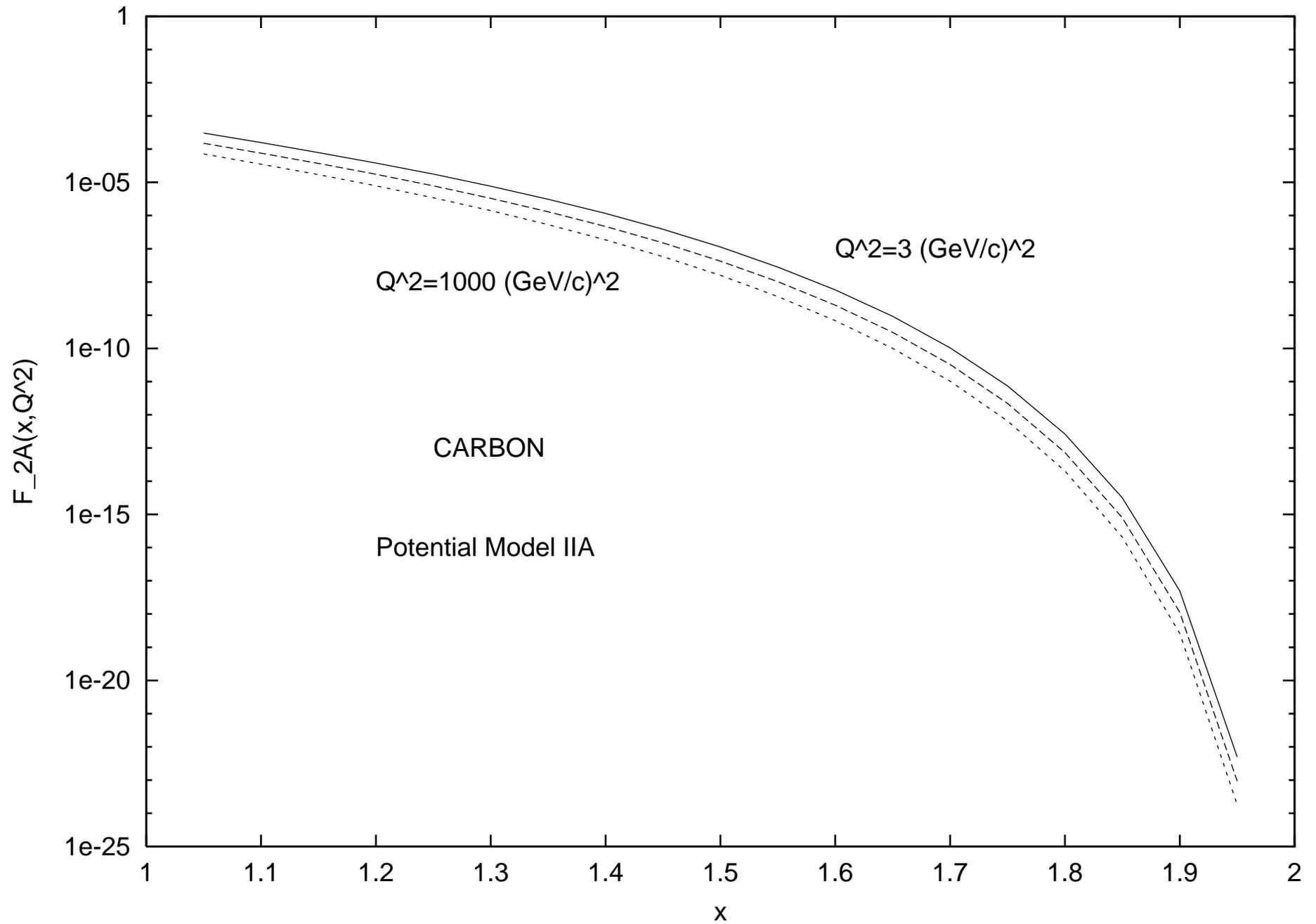
\end{document}